\documentclass[prd,twocolumn,preprintnumbers,showpacs,amsmath]{revtex4}

\usepackage{epsfig,amssymb,amsfonts,verbatim}

\def\bfl{\begin{flushleft}}
\def\efl{\end{flushleft}}
\def\bfr{\begin{flushright}}
\def\efr{\end{flushright}}
\def\bc{\begin{center}}
\def\ec{\end{center}}
\def\be{\begin{equation}}
\def\ee{\end{equation}}
\def\ba{\begin{eqnarray}}
\def\ea{\end{eqnarray}}
\def\baa#1{\begin{array}{#1}}
\def\eaa{\end{array}}
\def\bw{\begin{widetext}}
\def\ew{\end{widetext}}

\def\drm{d}

\def\schw{Schwarzschild  }

\def\lan{{\cal L}}

\newtheorem{Def}{Def.}[section]

\newtheorem{Prp}[Def]{Proposition}

\def\g{\gamma}

\def\k{\kappa}

\def\l{\lambda}

\def\m{\mu}

\begin{document}

\preprint{Phys. Rev. Lett. 94 (2005) 121101 ~~[hep-th/0408163]}

\title{
Coexistence of black holes and a long-range scalar field in cosmology
}

\author{Konstantin G. Zloshchastiev}
\affiliation{Instituto de Ciencias Nucleares, 
Universidad Nacional Aut\'onoma de M\'exico, A.P. 70-543,
 M\'exico D.F. 04510, M\'exico}




\begin{abstract}
The exactly solvable scalar hairy black hole model (originated from the
modern high-energy theory) is proposed.
It turns out that the existence of black holes (BH) is strongly correlated to 
global scalar field, in
a sense that they mutually impose bounds upon their physical 
parameters like the BH mass (lower bound) or the cosmological constant (upper bound).
We consider the same model also as a cosmological one 
and show that it agrees with recent experimental data; 
additionally, it provides a unified quintessence-like description of dark 
energy and dark matter.
\end{abstract}

\maketitle



\newpage

Two of the most fundamental predictions of the modern high-energy theory 
and gravity are black holes and cosmological scalar field (SF).
However, 
if existence of BH's has been experimentally confirmed since 70's
(and we even know now that BH's exist in centers of many galaxies including ours) 
then the global SF still lacks for direct experimental evidence,
mostly due to its extremely weak interaction with other matter.
In view of this, here we demonstrate that the good 
way to proceed would be to search for an influence the SF exerts 
on the first of the two phenomena we are considering here, black holes.

If one expects that the global ubiquitous scalar field does exist 
such that everything,
including black holes, is ``floating'' inside it, then one must allow the field
to get arbitrarily
close to the  surface of \textit{every} BH
which exists in the Universe at this moment. 
Moreover, to keep things physically consistent, when constructing models 
one must require that SF 
must be regular in the arbitrarily close vicinity of the event horizon.
This requirement, 
for a first look so innocent, in practice gave rise to
enormous technical difficulties.
In fact, beginning from 60's and until recently, nobody has succeeded in
satisfying it, i.e., in
finding the regular configurations of non-charged black holes and SF, 
the so-called scalar black holes (SBH).
By the latter one assumes the solution which is: 
(i) possessing an event horizon, 
(ii) physically acceptable (i.e., both the spacetime and SF
must be regular on and outside the horizon, have standard spherical topology and 
finite physical characteristics like mass, energy density, etc.; also the non-minimal
coupling, if any, must 
obey the recent 
observational bounds \cite{Will:2001mx}),
and 
(iii) not reducible to any other BH existing in absence of SF.
All these requirements have not yet
been fulfilled, despite the tremendous efforts and certain 
encouraging results \cite{Bocharova:1970,Bekenstein:1974sf}, 
see Ref. \cite{Martinez:2004nb}
for the most recent state of the art.
The models proposed so far either have unphysical features, like irregularities or 
exotic topology, or they involve additional
gauge fields, and then it becomes not clear why all BH's should have non-compensated
gauge charges to be consistent with global SF.
Even the numerical results are rare \cite{Torii:2001pg,Nucamendi:2003ex,Hertog:2004dr}.
Not without a sense of irony, people happened to be 
much more successful in solving the opposite
task: finding the requirements under which the physically admissible SBH can not exist, 
known as the scalar ``no-hair'' theorems \cite{Bizon:1994dh}, originated from 
the Wheeler's conjecture that
a BH can not be characterized by  
any parameters other than mass, electric charge and angular moment \cite{Ruffini:1971}. 
On the contrary, here we are going to solve this long-standing problem: 
we present the model which completely
satisfies the above-mentioned SBH criteria and thus ultimately falsifies the conjecture.

\textit{The model.}
We use the units where $16 \pi G = c = 1$, where $G$ is the
Newton  constant,
and
consider  theory describing  
self-interacting scalar field $\phi$ minimally coupled to Einstein
gravity.
Its Lagrangian is
$
\lan \sim 
R - (1/2)(\partial \phi)^2 - V (\phi) 
,
$
where 
the SF potential is given by
\be
V = 
2 \lambda (\cosh \phi + 2)
+ 4 \chi (3 \sinh \phi - \phi (\cosh \phi + 2))
,
\label{eV}
\ee
where $\chi$ and $\lambda$ are parameters of the model, 
$\lambda$ is usually called the cosmological constant. 
The model has the static spherically symmetric solution  
given in static observer coordinates 
$\{t,r,\theta,\varphi\}$ by
\be
\drm s^2 = - N^2  \drm t^2 + \frac{\drm r^2}{N^{2}}    + 
R^2 (\drm \theta^2 + \sin^2\!\theta \, \drm \varphi^2), \
e^\phi = H, 
\label{eSol}
\ee
where 
$H = 1 + \kappa/r^{\tilde d}$ is a harmonic function,
with $\tilde{d} = 1$ in our case,
$
N^2 = 1 - 2 \chi 
\left[
\kappa (r + \kappa/2) - R^2 
\ln H
\right]
-
\lambda R^2
$ with $R=\sqrt{r (r + \kappa)}$ being the habitual  radius
and $\k$ being the integration constant. 
The solution was obtained using the 
separability approach \cite{Zloshchastiev:2001da,Zloshchastiev:2001mw}.
The model also admits another solution which can be deduced
from (\ref{eSol}) by the simultaneous  inversion 
$\{\phi \to - \phi, \chi \to - \chi \}$ because our initial Lagrangian
has such $Z_2$ symmetry.
In other words, these two solutions can be grouped into a sort of duplet 
whose components are characterized 
by the discrete ``charge'' $Q \equiv \phi/|\phi|=\pm 1$.
For brevity, here we will work only with the solution 
we started from, corresponding to $Q=1$.
To clear its physical meaning, let us expand 
$N^2$ in series assuming $R/\k \propto r/\k \gg 1$ that gives the Newtonian limit.
We get 
$
N^2 \propto 1 - \l R^2 - \frac{2 M}{R} + \frac{\chi \k^5}{40 R^3} + O (1/R^5) 
,
$
where we have identified $M = \chi \k^3 / 6$ as a gravitating mass of (\ref{eSol}).
Thus, depending on whether $\l$ is zero, positive or negative, our solution
is asymptotically flat, de Sitter (dS) or anti-de Sitter (AdS).
Note, that the solution can not be reduced to the \schw one as the limit 
where SF vanishes corresponds to the de Sitter spacetime.

\begin{figure}[htbt]
\begin{center}\epsfig{figure=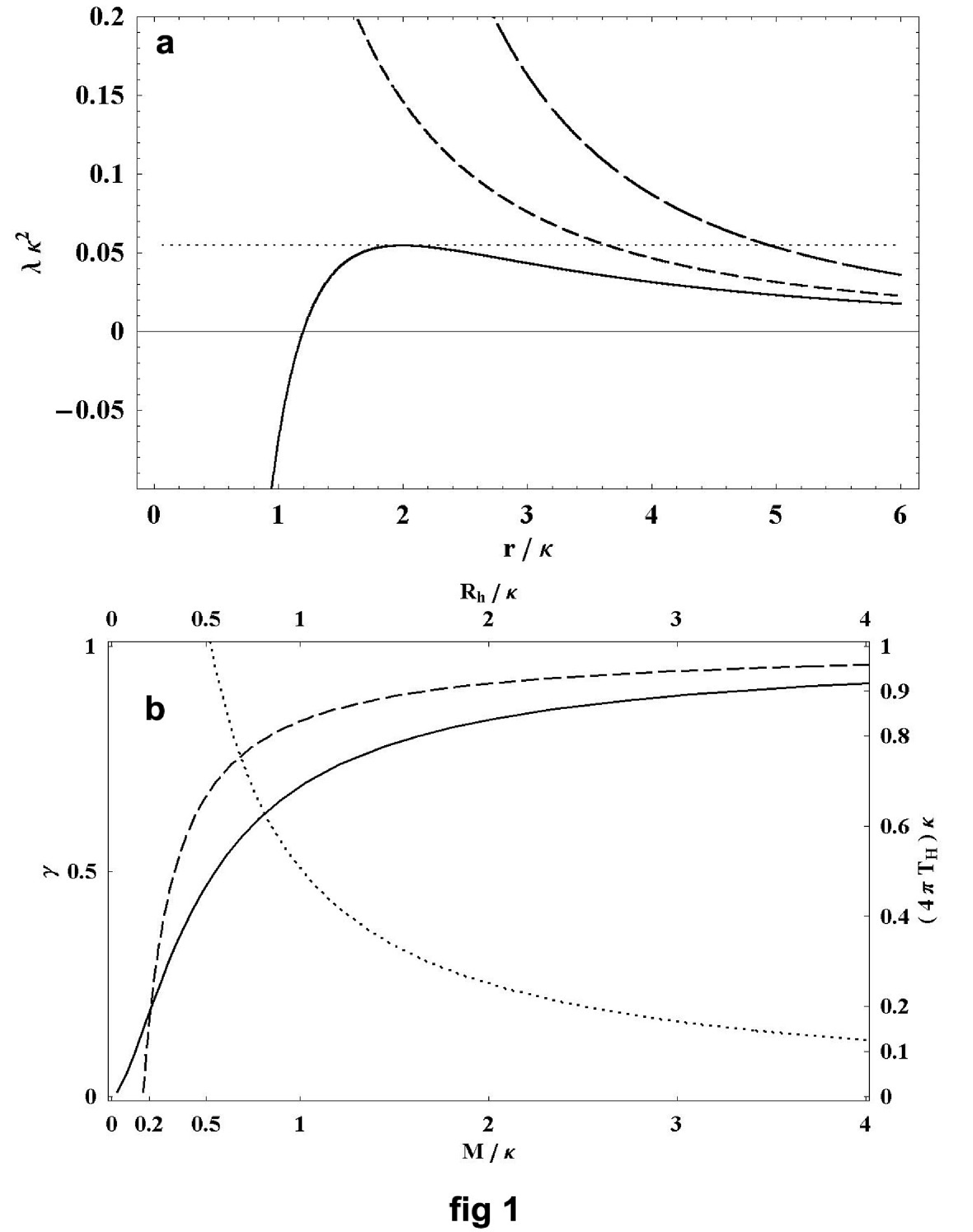,width=  0.9\columnwidth}\end{center}
\caption{
(a) Graphical solution of the horizon equation.
The intersections of curves with horizontal lines $\l \k^2$
yield the radii of horizons.
Cases: 
$\g \leqslant 0$ (short-dashed curve, plotted at $\g = -0.1$);
$\g > 1$ (long-dashed curve, plotted at $\g = 1.1$);
$0 < \g < 1$ (solid curve, plotted at $\g = 0.8$) where
the dotted horizontal line is $\l = \l_c$.
In all cases we have at least one horizon (the cosmological one)
but only in the third case we have the second, event, horizon.
(b) Radius, mass, temperature (here we assume $\l$ being negligibly small).
The solid and dashed curves correspond, respectively, to the radius $R_h$ 
and mass $M$ as functions of $\g$,
the dotted curve shows Hawking temperature $T_H$ as a function of $M$.
}
\label{f:1}
\end{figure}

Further, both SF and curvature invariants become singular at $R_s =0$, 
therefore, we 
may have problems with the Cosmic Censor unless the spacetime has the event
horizon located somewhere at 
$R_h = \sqrt{r_h (r_h + \kappa)} > R_s$ thus ``dressing'' the (otherwise naked) singularity.
The horizon condition is $N^2(r_h) = 0$, and its graphical
solution is shown in Fig. 1a where we have defined 
$\g \equiv 1 - (\chi \kappa^2)^{-1} \not= 1$.
From there one can deduce a few important things.
First, (\ref{eSol}) does describe SBH 
though not for every $\chi$ and $\k$ but only for those obeying
the inequality 
\be
0 < \g < 1
.
\label{eGamBnd}
\ee
Second, there exists an upper bound for $\l$: cosmological constant
must be below certain critical value $\l_c$ otherwise no black hole can exist.
Moreover, its absolute value must be much smaller than $\l_c$ 
to have the radius of a black hole 
much less than the size of the observable Universe:
$
\l \ll
\l_c \equiv 
4 \k^{-2}
\left(
\frac{\ln (2/\g - 1)
     }{2 (1-\g)}
- 1
\right)
.
$
Rough estimates give
$\l/\l_c \sim R_{\text{BH}}^2/R_{\text{Univ}}^2 \lesssim 10^{-28}$
if 
for maximal value of $R_{\text{BH}}$
we take that of the central-galactic BH's having mass 
$\lesssim 10^9 M_\odot$.
Thus, we see that the global parameter, cosmological
constant, turns out to be correlated with  
local quantities like the size of SBH or its mass.
Third, from ({\ref{eGamBnd}}) 
one can directly derive
that the previously defined mass of SBH $M$ is bounded from below:
$
M \geqslant M_{\text{extr}} \equiv M|_{\g \to + 0} = \k / 6
,
$
also $\g$ can be rewritten as $1 - M_{\text{extr}}/M$.
This property drastically differs SBH from the common BH solutions
like the \schw one where the lower bound is zero.
Fourth, ({\ref{eGamBnd}}) also gives the bounds for 
$\chi$ and $\k$: 
$\chi \geqslant \k^{-2} > 0$. 
The joint plot of the most important local
characteristics of our SBH (horizon radius, mass and Hawking temperature) is given
in Fig. 1b.
One can clearly see that when the horizon radius $R_h$ approaches zero the mass 
takes a non-zero value.

Finally, one should not forget that in the presence of SF the \schw solution can not
be regarded as a realistic one because the true vacuum (with 
vanishing stress-energy tensor $T_{\mu\nu}$) 
must be replaced by the SF background with
$
T_{\mu\nu} = 
\partial_\m \phi \, \partial_\nu \phi - 
g_{\mu\nu} 
\left[(1/2) (\partial \phi)^2 + V(\phi)
\right]
$.
In other words, the solutions like (\ref{eSol}) should be regarded
as describing the actual exterior gravitational field of massive bodies in ``vacuum''.

\begin{figure}[htbt]
\begin{center}
\epsfig{figure=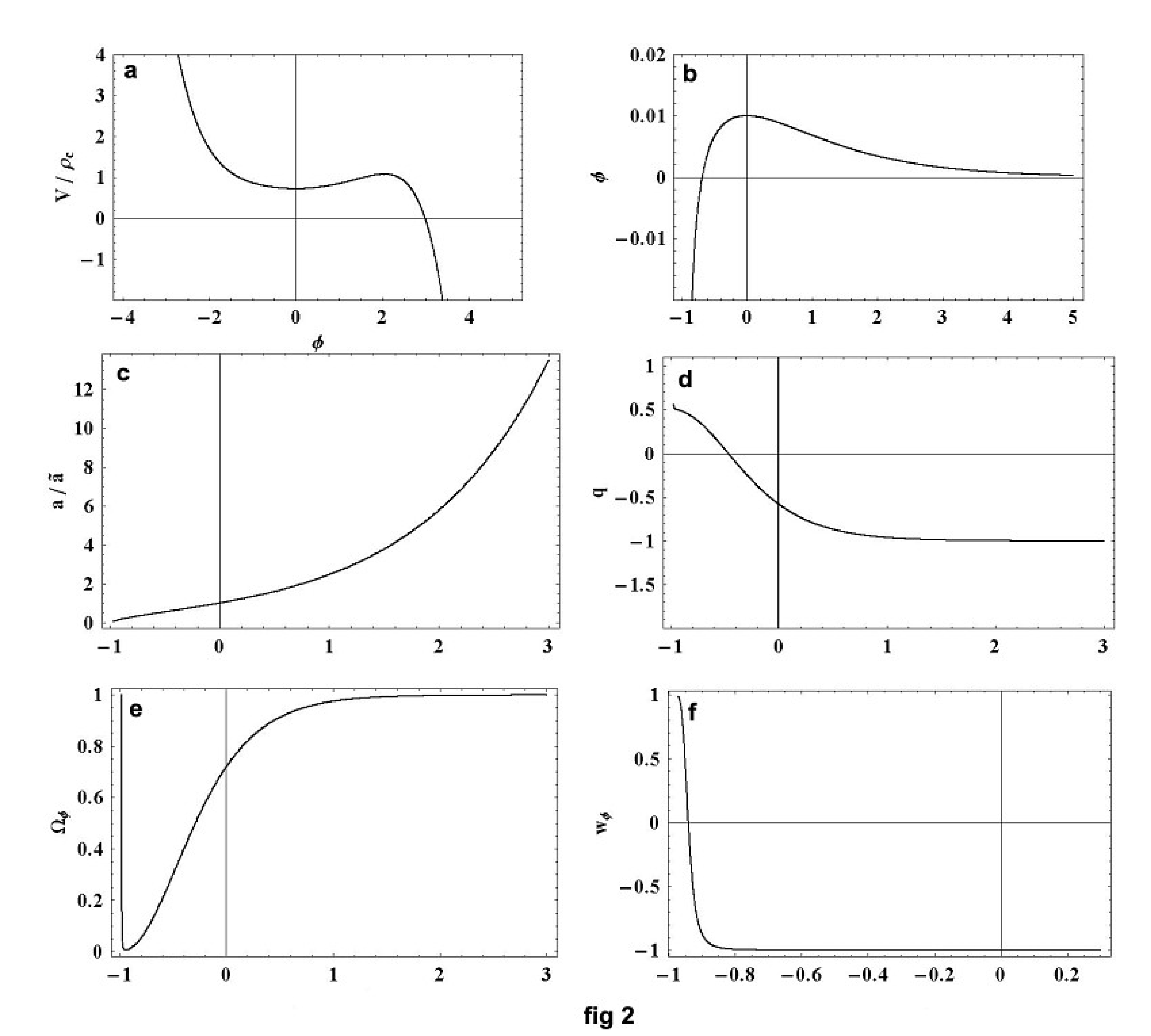,width=  \columnwidth}
\end{center}
\caption{
Cosmological scenario A.
For all plots except \textbf{a} 
the horizontal axis is the time measured in units $H_0^{-1} \sim 15$ billions years, and zero
corresponds to today.
The numerical input data: ID
$
\equiv
$
$
\{
\l/\rho_c,
\, \chi/\rho_c,
\, \phi|_{\tau=0}, 
\, (\drm \phi/\drm \tau)_{\tau=0}
\}
$
$=$
$\{ 0.12,\, 0.12,\, 10^{-2},\, 0 \}$,
where $\rho_c \sim 10^{-29} \, \text{g cm}^{-3}$ 
is the critical density.
We plot: (b) evolution of SF, 
(c) evolution of the size of the Universe,
(d) the deceleration parameter 
$q = - a\, \ddot a / \dot a^2$, 
(e) the dark energy ratio 
$\rho_\phi/\rho_{\text{tot}}$,
(f) the effective equation of state for SF
$p_\phi/\rho_\phi$.
}
\label{f:2}
\end{figure}

\textit{BH-compatible cosmology.} 
As long as we assume our SF being global and fundamental 
we must study cosmological consequences the model ({\ref{eV}}) implies.
The potential ({\ref{eV}}) consists of 
the symmetric and antisymmetric parts with respect to inversion of $\phi$,
proportional to $\l$ and $\chi$, respectively.
At small $\phi$ and non-zero cosmological constant $\l$
the symmetric part dominates: 
$V (\phi\to 0) \propto 6 \l (1+\phi^2/6) + O (\phi^4)$,
whereas for large values of the field it is the antisymmetric part that brings
the main contribution:
$V (\phi\to \pm \infty) \propto - 2 \chi \phi \exp |\phi|$.
Following the standard procedure adopted in cosmology we consider SF  
as a homogenous and isotropic function of cosmological time, $\phi(t)$, 
and conduct numerical
simulations for our model at different values of its parameters.
They showed that the following scenarios of the spatially flat 
FRW Universe evolution are possible.

In the mainstream cosmological scenario with positive $\l$,
our SF (inflaton) starts at $\phi \ll - 1$, rolls down towards 
the local dS minimum of the potential, passes it and tries to climb over the local
dS maximum. If its initial energy is not sufficient to do that 
(note that the inflaton's motion is not ``frictionless'':
there exists sufficient dissipation of energy for creation of radiation
and matter) then we have the Scenario A: 
inflaton rolls back to the local minimum asymptotically
approaching the value $\phi = 0$, as in Fig. 2b.
Meanwhile, the Universe experiences an accelerated expansion, see Fig. 2b,
with the eternal acceleration, Fig. 2c. 
The ratio of the SF density (dark energy) to the
total energy density approaches the approximate value $0.72$, Fig. 2d.
Figure 2f reveals the quintessential-like \cite{Zlatev:1998tr}
behavior of SF: 
during some epoch in past (or, equivalently, in ``redshifted'' regions)
it behaved like a pressureless matter
($w_\phi \equiv p_\phi / \rho_\phi \sim 0$) but afterwards
its effective equation of state became of the false vacuum type 
($w_\phi \to w_{\text{DE}} = -1$).
Thus, the model provides a unified description
of dark energy and dark matter without \textit{ad hoc} assumptions -
they appear to be different manifestations of the same entity,
scalar field.
All these results agree with the recent experimental 
data coming from high-redshift observations of supernovae 
\cite{Garnavich:1997nb,Riess:1998cb,Perlmutter:1998np} and 
anistropies of the cosmic microwave background (CMB) spectrum
\cite{Dodelson:1999am,Netterfield:2001yq,Stompor:2001xf,Halverson:2001yy}.

\begin{figure}[htbt]
\begin{center}
\epsfig{figure=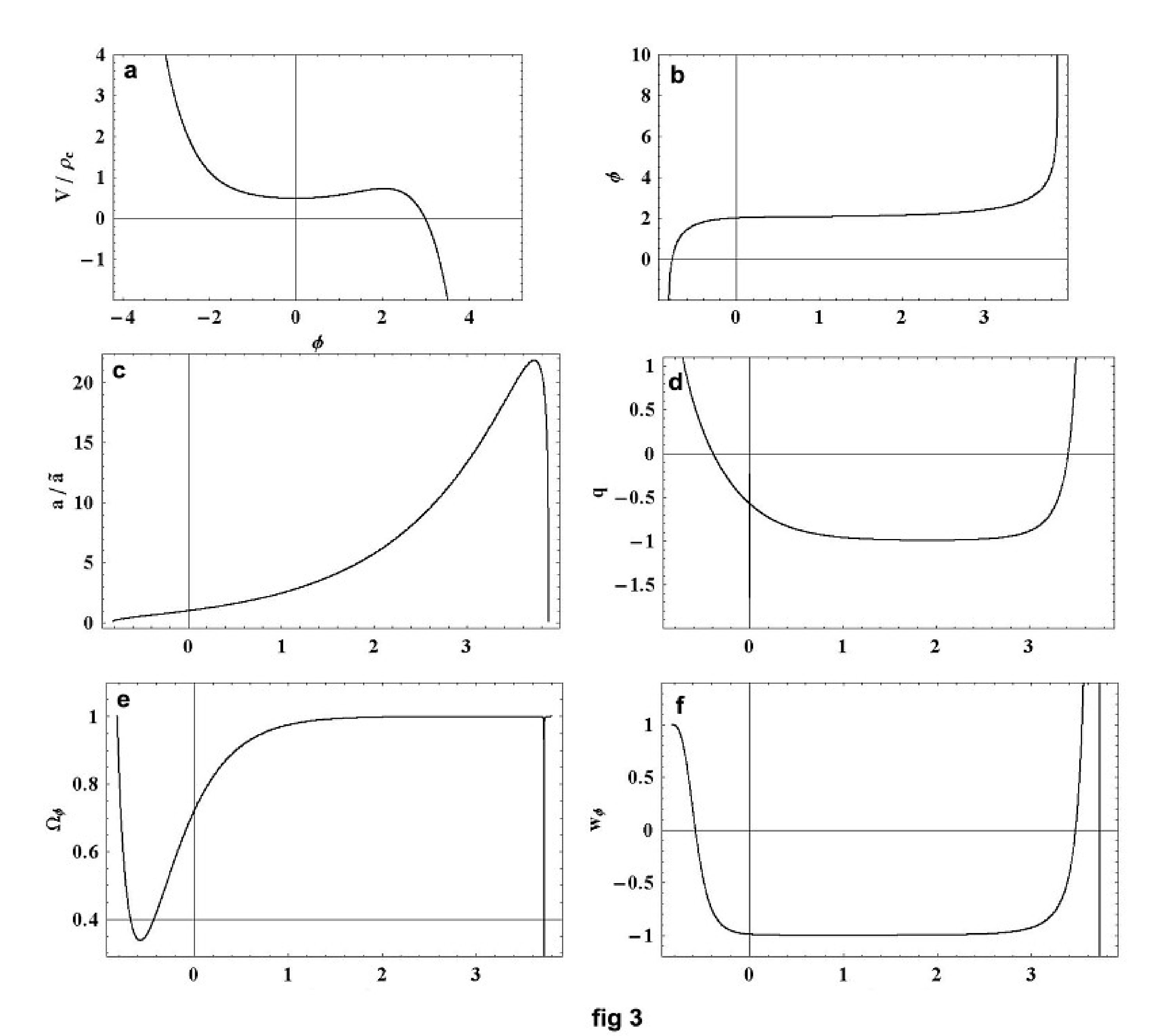,width=  \columnwidth}
\end{center}
\caption{
Cosmological scenario B.
Notations are the same as in previous Fig.;
ID
$=$
$
\{ 0.08,\, 0.08,\, 2,\, 0.2 \}
$.
}
\label{f:3}
\end{figure}

In another case, 
when the initial energy of SF is sufficient to overcome the local
maximum (classically or by virtue of the tunnelling effect), 
we have the Scenario B: scalar field rolls over the maximum 
and starts moving all the way down (Fig. 3b),
whereas the Universe at some point begins to decelerate (Fig. 3d), and eventually
collapses to the Big Crunch (Fig. 3c), even despite being spatially flat. 
Figure 3e shows that this scenario is in agreement 
with experiment too. 
Thus, one may wonder which scenario is the real one, A (ever-expanding Universe) or B
(Big Crunch)? 
At the classical level - we do not know as we do not know the recent value of SF and its rate of change
which are related to the initial ones.
The taking into account quantum tunnelling makes Big Crunch 
an inevitable final state, as will be discussed below.

In principle, the presented scenarios 
are sufficient
to show the viability of the cosmological model based on Eq. ({\ref{eV}}).
Yet, we would like to consider also other possibility -  
when $\l$ is not necessarily positive.
The reason is that many people associate the accelerated expansion of the Universe
with a positive cosmological constant and the regime where scalar field
approaches the dS vacuum state.
Let us demonstrate how this stereotype gets broken in our model.
By numerical simulations one can easily show that 
the accelerated expansion of the Universe may occur not only when 
$\l > 0$ but also at $\l \leqslant 0$, when no dS extrema exist.
Let us consider the case $\l = 0$ only, because the case of a negative cosmological
constant (AdS) is very similar qualitatively.
In this case the potential has a saddle point at $\phi=0$ instead of
two local extrema, see Fig. 4a.
One can imagine the following two scenarios.
First takes place
when  the initial value of SF is large negative such that
initially it ``sits uphill'' and starts unbounded motion all the way down,
as time goes.
Second one happens when initially $\phi$ is situated ``downhill'' 
(such that it is large positive)
but its initial kinetic energy is large enough to climb up.
Then it moves up, passes the saddle point, continues an ascent until
reaches the maximum point
of its trajectory, and then rolls back all the way down, see Fig. 4 b.
The accelerated expansion of the Universe takes place in both cases.
However, in the first case the inflation ends too soon such that one could not
detect any acceleration nowadays, 
neither the dark energy approaches its experimentally established value today. 
The second scenario is better in this connection: 
the Universe passes a certain epoch
of accelerated expansion whose time can be tuned to coincide 
with today, Fig. 4d. 
Thereby the size of the Universe evolves with time as in Fig. 4c: 
decelerated expansion,
accelerated expansion, again decelerated expansion, shrinking, and finally, Big Crunch.
The recent value of $\Omega_\phi$ agrees with experiment, Fig. 4e.

To summarize, 
our BH-compatible cosmological model ({\ref{eV}}) seems to be compatible with the
experimental data 
for large range of its parameters.
Besides, above
our model 
has been \textit{explicitly} proven to 
be consistent with the existence of black holes in the Universe.
The class of such models can not be vast \textit{a priori} because
the scalar ``no-hair'' theorems forbid the appearance of black holes for a 
large set of the SF potentials, 
e.g., convex or positive semi-definite 
\cite{Bekenstein:1972ny,Sudarsky:1995zg}.
Thus, 
the existence of BH's can be the strong criterion for 
theoretical cosmology sufficiently narrowing the 
class of physically admissible models. 

\begin{figure}[htbt]
\begin{center}
\epsfig{figure=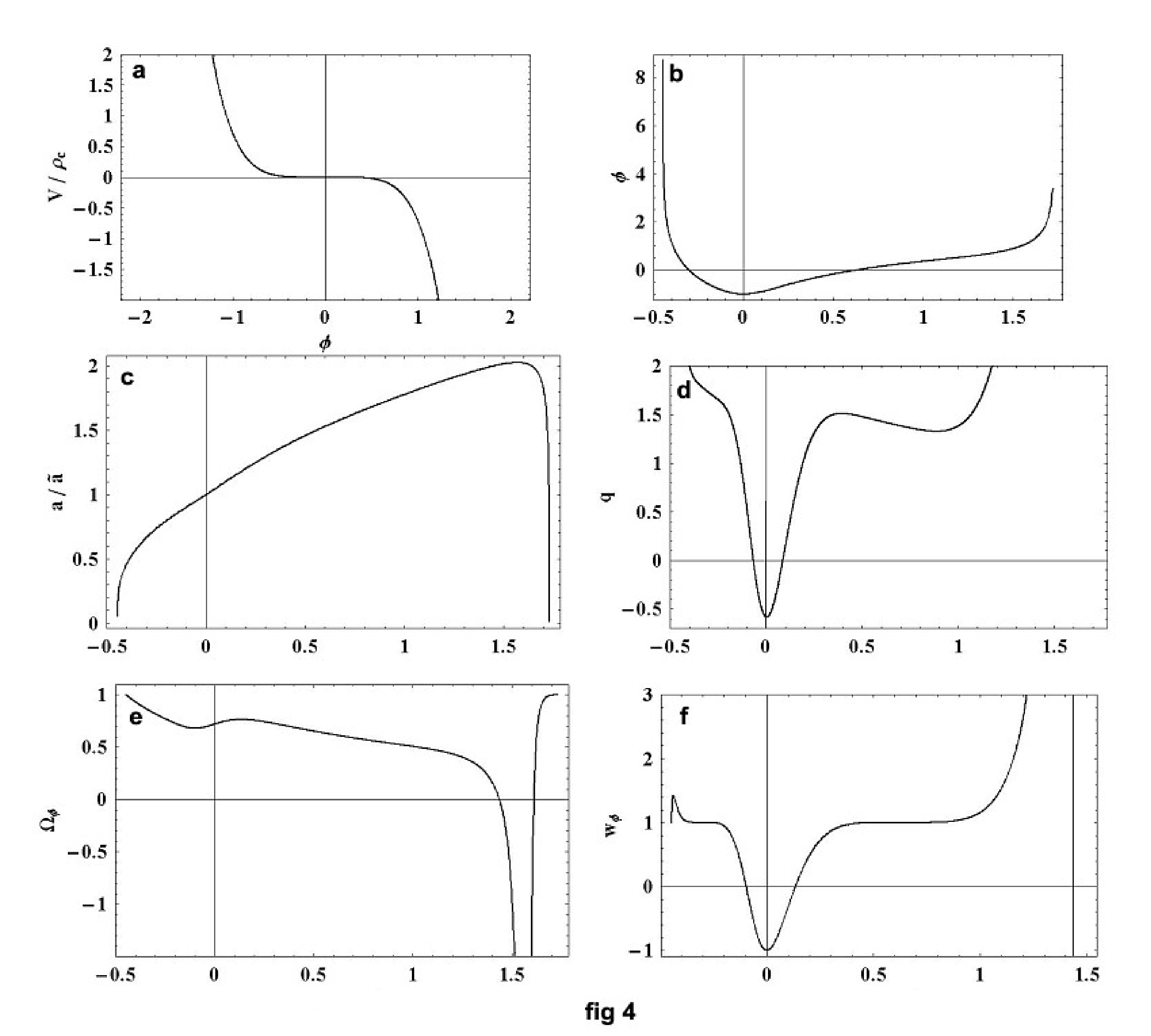,width=  1\columnwidth}
\end{center}
\caption{
Scenario with $\l = 0$.
ID
$=$
$
\{ 0,\, 10.3,\, -1, 0 \}
$.
}
\label{f:4}
\end{figure}

\textit{Origin of the model.}
There is nothing wrong in regarding our model as a postulated
one, yet the picture would not be complete without mentioning where
({\ref{eV}}) came from. 
It appears in the novel class of the four-dimensional (4D) effective field theories 
(EFT) which describe the low-energy limit of
the 11D M-theory taking into account
the non-perturbative aspects such as BPS states and 
$p$-branes \cite{Zlosh}.
The scalar potential in those EFT's in the simplest case satisfies 
the second order differential 
equation: 
$
V'' - (a + a^{-1}) 
\coth 
\left[ (a + a^{-1}) \phi/4
\right] V' + V = 0
,
$
where the constant $a$ is precisely the one that appears 
in the truncated supergravity (SUGRA), 
$\lan \sim R - (1/2) (\partial \phi)^2 - (1/2n !)\text{e}^{a \phi} F^2_{[n]} + ...$ .
The potential ({\ref{eV}}) arises as a solution of this ODE at $a^2=1$
(other potentially supersymmetric cases $a^2 = 3,\, 1/3$
also have been studied by author but are not listed here).
The considering those EFT's goes far beyond the scope 
of this paper, instead we just briefly outline 
some common features of our model and SUGRAs.
For instance, the structure of the solution ({\ref{eSol}}) 
looks very similar to that of $0$-branes \cite{Duff:1996hp,Stelle:1998xg}.
The Breitenlohner-Freedman parameter \cite{Breitenlohner:1982bm}
takes the conformal value 
$
\m^2 
= -2 > - 9/4
$
thus the model's stability can be enhanced by partially unbroken supersymmetry.
Further, the symmetric $\l$-part of our potential   
resembles those arising in SUGRA-inspired cosmologies \cite{Hertog:2004dr,DL:1999,Kallosh:2001gr},
however, our model also has the antisymmetric $\chi$ part which is responsible for 
black holes (and at the same time for cosmological behavior at large values of SF).
Besides, the $\chi$ part is 
appreciated by string theory and QFT in curved spacetime
because of certain 
conceptual difficulties 
with the ever-accelerating dS Universe \cite{Sasaki:1992ux,Banks:2001yp}.
From the quantum viewpoint, even in the Scenario $A$ the local (dS) minimum is a quasi-bound state 
and thus the system can stay there only for a finite time -
eventually it  tunnels through the local maximum, 
such that its further dynamics will be as in Scenario $B$.

\begin{acknowledgments} 
I thank D. Sudarsky, A. G\"uijosa, M. Salgado and C. Chryssomalakos for
enlightening discussions. 
\end{acknowledgments}

\def\AnP{Ann. Phys.}
\def\APP{Acta Phys. Polon.}
\def\CJP{Czech. J. Phys.}
\def\CMPh{Commun. Math. Phys.}
\def\CQG {Class. Quantum Grav.}
\def\EPL  {Europhys. Lett.}
\def\IJMP  {Int. J. Mod. Phys.}
\def\JMP{J. Math. Phys.}
\def\JPh{J. Phys.}
\def\FP{Fortschr. Phys.}
\def\GRG {Gen. Relativ. Gravit.}
\def\GC {Gravit. Cosmol.}
\def\LMPh {Lett. Math. Phys.}
\def\MPL  {Mod. Phys. Lett.}
\def\Nat {Nature}
\def\NCim {Nuovo Cimento}
\def\NPh  {Nucl. Phys.}
\def\PhE  {Phys.Essays}
\def\PhL  {Phys. Lett.}
\def\PhR  {Phys. Rev.}
\def\PhRL {Phys. Rev. Lett.}
\def\PhRp {Phys. Rept.}
\def\RMP  {Rev. Mod. Phys.}
\def\TMF {Teor. Mat. Fiz.}
\def\prp {report}
\def\Prp {Report}

\def\jn#1#2#3#4#5{{#1}{#2} {\bf #3}, {#4} {(#5)}} 

\def\boo#1#2#3#4#5{{\it #1} ({#2}, {#3}, {#4}){#5}}



\end{document}